%% file: Duplij-Vogl_Innov-squbits.tex
\documentclass[12pt,reqno,twoside]{amsart}%
\usepackage{amsmath,amsthm,amscd,amsthm,upref,indentfirst}
\usepackage{amsfonts,mathrsfs}

\usepackage{amssymb,amsbsy,bm}
\usepackage{graphicx}
\usepackage{times}
\usepackage{amsaddr}

\usepackage{soul}
\usepackage{amsxtra}
\usepackage{latexsym}
\usepackage{mathrsfs}
\usepackage{amsthm,upref,indentfirst}
\usepackage{amsbsy}
\usepackage{mathdots}
\usepackage{mathabx}

\usepackage{mathabx}

\usepackage[usenames,dvipsnames]{color}

\theoremstyle{plain}
\newtheorem{theorem}{Theorem}

\theoremstyle{definition}
\newtheorem{definition}[theorem]{Definition}

\theoremstyle{remark}
\newtheorem{remark}[theorem]{Remark}

\numberwithin{equation}{section}
\numberwithin{theorem}{section}

\usepackage{exscale}
\usepackage[scaled=0.86]{helvet}

\normalfont\upshape

\usepackage[scr=euler,scrscaled=1.1,bb=txof,bbscaled=1.16,cal=boondox,calscaled=1.1]{mathalfa}
\renewcommand{\mathit}{\bm}
\renewcommand{\mathfrak}{\mathscr}

\renewcommand{\mathtt}[1]{\scalebox{1.2}{\bf \texttt{\upshape#1}}}
\renewcommand{\emph}[1]{\textcolor{blue}{\textbf{#1}}}

\usepackage{float}

\usepackage[justification=centering]{caption}

\numberwithin{equation}{section}
\numberwithin{theorem}{section}

\usepackage[normalem]{ulem}
\usepackage[square,comma]{natbib}

\usepackage{mparhack}

\usepackage{keyval}
\usepackage{totcount}
\newtotcounter{citnum} 
\def\oldbibitem{} \let\oldbibitem=\bibitem
\def\bibitem{\stepcounter{citnum}\oldbibitem}

\usepackage[verbose]{backref}
\backrefsetup{verbose=false}
\renewcommand*{\backref}[1]{}
\renewcommand*{\backrefalt}[4]{[{\tiny%
    \ifcase #1 \textsl{Not cited}%
          \or \textsl{Cited on page}~\textcolor{BrickRed}{#2}%
          \else \textsl{Cited on pages}~\textcolor{BrickRed}{#2}%
    \fi%
    }]}

\usepackage{fancyhdr}
\author{\small\scshape S\lowercase{teven} D\lowercase{uplij} \lowercase{and} R\lowercase{aimund} V\lowercase{ogl}}

\address{
Center for Information Technology,
Universit\"at M\"unster,
R\"ontgenstrasse 7-13\\
D-48149 M\"unster,
Deutschland}
\email{\small \sf douplii@uni-muenster.de;
sduplij@gmail.com;
https://ivv5hpp.uni-muenster.de/u/douplii}
\title{\large\bfseries\scshape
O\lowercase{n superqubits}}

\date{\textit{of start} September 22, 2023. \textit{Date}:
\textit{of completion}
October 14, 2023.
\newline
\mbox{}\hskip 1.16em
\textit{Total}:
35
references.%
}

\renewcommand{\refname}{\textsc{References}}

\let\origsection\section
\renewcommand{\section}[1]{\sectionmark{#1}\origsection{#1}}
\let\origsubsection\subsection
\renewcommand{\subsection}[1]{\subsectionmark{#1}\origsubsection{#1}}

\makeatletter
\renewenvironment{thebibliography}[1]{%
  \@xp\origsection\@xp*\@xp{\refname}%
  \normalfont\footnotesize\labelsep .9em\relax
  \renewcommand\theenumiv{\arabic{enumiv}}\let\p@enumiv\@empty
  \vspace*{-5pt}
  \list{\@biblabel{\theenumiv}}{\settowidth\labelwidth{\@biblabel{#1}}%
    \leftmargin\labelwidth \advance\leftmargin\labelsep
    \usecounter{enumiv}}%
  \sloppy \clubpenalty\@M \widowpenalty\clubpenalty
  \sfcode`\.=\@m
}{%
  \def\@noitemerr{\@latex@warning{Empty `thebibliography' environment}}%
  \endlist
}
\makeatother

\subjclass[2010]{17A40, 17A42, 20N10, 20N15, 58A50, 81P40, 81P45, 81P68}
\keywords{supermatrix, even, odd, super Hilbert space, concurrence, qubit, qutrit, superqubit}

\begin{document}
\mbox{}
\vspace{1cm}

\mbox{}
\begin{abstract}
\input{
Duplij-Vogl_Innov-squbits-abs.tex}
\end{abstract}
\maketitle

\thispagestyle{empty}


\mbox{}
\vspace{1.5cm}
\tableofcontents
\newpage

\pagestyle{fancy}

\addtolength{\footskip}{15pt}

\renewcommand{\sectionmark}[1]{%
\markboth{
{ \scshape #1}}{}}

\renewcommand{\subsectionmark}[1]{%
\markright{
\mbox{\;}\\[5pt]
\textmd{#1}}{}}

\fancyhead{}
\fancyhead[EL,OR]{\leftmark}
\fancyhead[ER,OL]{\rightmark}
\fancyfoot[C]{\scshape -- \textcolor{BrickRed}{\thepage} --}

\renewcommand\headrulewidth{0.5pt}
\fancypagestyle {plain1}{ %
\fancyhf{}
\renewcommand {\headrulewidth }{0pt}
\renewcommand {\footrulewidth }{0pt}
}

\fancypagestyle{plain}{ %
\fancyhf{}
\fancyhead[C]{\scshape S\lowercase{teven} D\lowercase{uplij} \hskip 0.7cm \MakeUppercase{Polyadic Hopf algebras and quantum groups}}
\fancyfoot[C]{\scshape - \thepage  -}
\renewcommand {\headrulewidth }{0pt}
\renewcommand {\footrulewidth }{0pt}
}

\fancypagestyle{fancyref}{ %
\fancyhf{} 
\fancyhead[C]{\scshape R\lowercase{eferences} }
\fancyfoot[C]{\scshape -- \textcolor{BrickRed}{\thepage} --}
\renewcommand {\headrulewidth }{0.5pt}
\renewcommand {\footrulewidth }{0pt}
}

\fancypagestyle{emptyf}{
\fancyhead{}
\fancyfoot[C]{\scshape -- \textcolor{BrickRed}{\thepage} --}
\renewcommand{\headrulewidth}{0pt}
}
\mbox{}
\thispagestyle{emptyf}
\input{Duplij-Vogl_Innov-squbits-sw}

\pagestyle{emptyf}

\input{Duplij-Vogl_Innov-squbits.bbl}
\end{document}

%% file: Duplij-Vogl_Innov-squbits-abs.tex

\noindent We first reconsider the mathematical background of superqubit theory
and describe important peculiarities of superspaces and supermatrices which
are usually out of attention. Then we study states in super Hilbert spaces
using super-bra/super-ket formalism in details. The qubit (qudit) and
superqubit (superqudit) are defined as linear spans in the corresponding
Hilbert subspaces. A new kind of superqubit carring the odd parity is
introduced. The multi-superqubit states are studied, and the superconcurrence
which distinguishes separable states is proposed.

%% file: Duplij-Vogl_Innov-squbits-sw.tex

\section{\textsc{Introduction}}

It is well known that the quantum computation is based on the algebraic
structure of it constituents, qubits and qudits, \textquotedblleft
living\textquotedblright\ in some Hilbert space. Therefore, possible
improvements could be connected with some special generalizations of the
Hilbert space. One of promising direction is supersymmetric generalization of
the ordinary Hilbert space \cite{dewitt,const2} and consideration of various
super analogs of quantum states in it with simultaneous passing from
corresponding groups to supergroups.

\section{\textsc{Superspaces and supermatrices}}

Let us consider the main ideas in supersymmetrization of qubits along
\cite{bor/dah/duf/rub,bor/bra/duf,bor/bra/duf1}. The principal statement is
changing the Hilbert space to super Hilbert space (in sense of \cite{rud2000})
and considering quantum states as (even) supervectors in the latter, i.e.
taking values in the corresponding Grassmann algebra (or some more general
supercommutative superalgebra). In this approach the inner product and
probabilities contain Grassmann algebra parts. In the same way the bra/ket
formalism of quantum mechanics \cite{dir1939,eij/gra,gie2000} transforms to
super-bra/super-ket one with additional parity rules. Here we will point out
the foremost relations and statements concisely (only needed), while refer to
the details and further notations to the standard supermathematics sources
\cite{berezin,lei1,dewitt}. To clarify the structure of variables we present
some formulas in two columns: ordinary (left) and supersymmetric (right)
cases, and moreover we use different notations for them (the latter will be
marked in bold).

Let $\Lambda_{N}\left(  \mathbb{C}\right)  $ be a complex Grassmann algebra
having $N$ anticommuting generators $\theta_{i}$. The nilpotence of $\theta$'s
(which follows from their anticommutativity) leads to its finiteness (with
dimension $2^{N}$) and to the decompositions of any element $\mathit{z}%
\in\Lambda_{N}\left(  \mathbb{C}\right)  $ (informally)%
\begin{equation}
\mathit{z}=\overset{\text{no\ }\theta\text{'s}}{\overbrace{z_{body}}}%
+\overset{\text{with }\theta\text{'s}}{\overbrace{\mathit{z}_{soul}}}%
=\overset{\text{no\ }\theta\text{'s and no }0}{\overbrace{z_{invert}}%
}+\overset{\text{with }\theta\text{'s and }0}{\overbrace{\mathit{z}%
_{noninvert}}=}\overset{\text{even\ }\theta\text{'s}}{\overbrace
{\mathit{z}_{even}}}+\overset{\text{odd }\theta\text{'s}}{\overbrace
{\mathit{z}_{odd}}}, \label{qs-z}%
\end{equation}
where $z_{body}\in\mathbb{C}$, $z_{invert}\in\mathbb{C}\setminus0$,
$\mathit{z}_{soul}\in\Lambda_{N}\left(  \mathbb{C}\right)  \setminus
\mathbb{C}$, $\mathit{z}_{noninvert}\in\Lambda_{N}\left(  \mathbb{C}\right)
\setminus\mathbb{C\cup}\left\{  0\right\}  $. The last decomposition allows us
to introduce the degree by $\deg\mathit{z}_{even}=\bar{0}$ and $\deg$
$\mathit{z}_{odd}=\bar{1}$, and elements with the fixed degree are
homogeneous. Obviously, that for homogeneous elements $\deg\mathit{yz}%
=\deg\mathit{y}+\deg\mathit{z}\left(  \operatorname{mod}2\right)  $. Another
name of $\deg$ is parity (or grade), in special cases they are fine different
\cite{leites13}, but in the superqubit context all of them are interchangable.
Thus, the mapping $\deg:\Lambda_{N}\left(  \mathbb{C}\right)  \rightarrow
\mathbb{Z}_{2}=\left\{  \bar{0},\bar{1}\right\}  $, gives the direct sum
decomposition of the Grassmann algebra $\Lambda_{N}\left(  \mathbb{C}\right)
=\Lambda_{N}^{\left(  even\right)  }\left(  \mathbb{C}\right)  \oplus
\Lambda_{N}^{\left(  odd\right)  }\left(  \mathbb{C}\right)  =\Lambda
_{N}^{\left(  \bar{0}\right)  }\left(  \mathbb{C}\right)  \oplus\Lambda
_{N}^{\left(  \bar{1}\right)  }\left(  \mathbb{C}\right)  $, which means that
$\Lambda_{N}\left(  \mathbb{C}\right)  $ is the simplest example of
$\mathbb{Z}_{2}$-graded algebra. The analog of the ordinary commutator for
$\Lambda_{N}\left(  \mathbb{C}\right)  $ is the supercommutator%
\begin{equation}%
\begin{tabular}
[c]{ccc}%
$\left[  y,z\right]  =yz-zy$, $\ y,z\in\mathbb{C}$, & $\overset{susy}%
{\Longrightarrow}$ & $\left[  \mathit{y},\mathit{z}\right]  _{\deg
}=\mathit{yz}-\left(  -1\right)  ^{\deg\mathit{y}\deg\mathit{z}}%
\mathit{zy},\ \mathit{y},\mathit{z}\in\Lambda_{N}\left(  \mathbb{C}\right)  .$%
\end{tabular}
\ \ \label{qs-yz}%
\end{equation}

If $\left[  \mathit{y},\mathit{z}\right]  _{\deg}=0$ for all elements of a
superalgebra, then it is supercommutative, which is indeed the case of the
Grassmann algebra $\Lambda_{N}\left(  \mathbb{C}\right)  $. The same rule will
be implied for all other $\mathbb{Z}_{2}$-graded homogeneous variables.

The ordinary involution $\ast$ and the grade involution $\sharp$ (superstar or
superinvolution \cite{leites13,bor/dah/duf/rub}) can be defined on
$\Lambda_{N}\left(  \mathbb{C}\right)  $ as follows%
\begin{align}
\left(  x\mathit{y}\right)  ^{\ast}  &  =\bar{x}\mathit{y},\ \ \left(
\mathit{yz}\right)  ^{\ast}=\mathit{z}^{\ast}\mathit{y}^{\ast},\ \ \left(
\mathit{y}\right)  ^{\ast\ast}=\mathit{y},\label{qs-star}\\
\left(  x\mathit{y}\right)  ^{\sharp}  &  =\bar{x}\mathit{y}^{\sharp
},\ \ \left(  \mathit{yz}\right)  ^{\sharp}=\mathit{y}^{\sharp}\mathit{z}%
^{\sharp},\ \ \left(  \mathit{y}\right)  ^{\sharp\sharp}=\left(  -1\right)
^{\deg\mathit{y}}\mathit{y},\ \ \ \ x\in\mathbb{C},\ \ \mathit{y}%
,\mathit{z}\in\Lambda_{N}\left(  \mathbb{C}\right)  , \label{qs-sstar}%
\end{align}
such that $\mathit{z}^{\ast}\in\Lambda_{N}^{\left(  \deg\mathit{z}\right)
}\left(  \mathbb{C}\right)  $,\ $\mathit{z}^{\sharp}\in\Lambda_{N}^{\left(
\deg\mathit{z}\right)  }\left(  \mathbb{C}\right)  $.

The superqubits \textquotedblleft live\textquotedblright\ in a
finite-dimensional $\mathbb{Z}_{2}$-graded linear vector space (or superspace)
$\mathcal{V}$ over $\mathbb{C}$ (or any other field $\mathbb{K}$) which has
the same decomposition on the even and odd parts as the Grassmann algebra
above $\mathcal{V}=\mathfrak{V}^{\left(  \bar{0}\right)  }\oplus
\mathfrak{V}^{\left(  \bar{1}\right)  }$. If the dimensions of the component
spaces $\dim\mathfrak{V}^{\left(  \bar{0}\right)  }=p$ and $\dim
\mathfrak{V}^{\left(  \bar{1}\right)  }=q$, then we denote the $\mathbb{Z}%
_{2}$-graded vector space $\mathcal{V}=\mathbb{C}^{p|q}$, and its
\textquotedblleft graded\textquotedblright\ dimension $\dim\mathbb{C}%
^{p|q}=p+q$. The $\mathbb{Z}_{2}$-graded direct product of superspaces
$\hat{\otimes}$ (which is used for superqubit constructions) is crucially
different from the ordinary direct product of spaces $\otimes$ (exploited for
qubits). Indeed, we have the ordinary direct product%
\begin{equation}
\mathcal{V}\otimes\mathcal{W}=\mathfrak{V}^{\left(  \bar{0}\right)  }%
\otimes\mathfrak{W}^{\left(  \bar{0}\right)  }+\mathfrak{V}^{\left(  \bar
{1}\right)  }\otimes\mathfrak{W}^{\left(  \bar{1}\right)  }+\mathfrak{V}%
^{\left(  \bar{1}\right)  }\otimes\mathfrak{W}^{\left(  \bar{0}\right)
}+\mathfrak{V}^{\left(  \bar{0}\right)  }\otimes\mathfrak{W}^{\left(  \bar
{1}\right)  }, \label{qs-vw0}%
\end{equation}
which does not allow us to introduce the $\mathbb{Z}_{2}$-graded structure
without additional assumptions.

Only the definition of a new operation, the $\mathbb{Z}_{2}$-graded direct
product $\hat{\otimes}$, gives the consistent superspace structure of product
(by endowing two last terms in (\ref{qs-vw0}) with odd grading of the product)%
\begin{equation}
\left(  \mathcal{V}\hat{\otimes}\mathcal{W}\right)  ^{\left(  \mathsf{k}%
\right)  }=%
{\displaystyle\bigoplus\limits_{\mathsf{k}=\mathsf{r}\boxplus\mathsf{s}}}
\mathfrak{V}^{\left(  \mathsf{r}\right)  }\otimes\mathfrak{W}^{\left(
\mathsf{s}\right)  },\ \ \ \ \ \mathsf{k},\mathsf{r},\mathsf{s}\in
\mathbb{Z}_{2},\ \ \ \mathsf{r}\boxplus\mathsf{s}\equiv\mathsf{r}%
+\mathsf{s}\left(  \operatorname{mod}2\right)  , \label{qs-vw}%
\end{equation}
or simply%
\begin{align}
\left(  \mathcal{V}\hat{\otimes}\mathcal{W}\right)  ^{\left(  \bar{0}\right)
}  &  =\mathfrak{V}^{\left(  \bar{0}\right)  }\otimes\mathfrak{W}^{\left(
\bar{0}\right)  }+\mathfrak{V}^{\left(  \bar{1}\right)  }\otimes
\mathfrak{W}^{\left(  \bar{1}\right)  },\\
\left(  \mathcal{V}\hat{\otimes}\mathcal{W}\right)  ^{\left(  \bar{1}\right)
}  &  =\mathfrak{V}^{\left(  \bar{1}\right)  }\otimes\mathfrak{W}^{\left(
\bar{0}\right)  }+\mathfrak{V}^{\left(  \bar{0}\right)  }\otimes
\mathfrak{W}^{\left(  \bar{1}\right)  }.
\end{align}
Usually, the (different) operations $\hat{\otimes}$ and $\otimes$ are denoted
by the same symbol, but they should be used with care and taking account the
actual distinction of (\ref{qs-vw0}) and (\ref{qs-vw}).

In the consideration of mappings between superspaces and trying to introduce
$\mathbb{Z}_{2}$-graded structure for them, we also note some peculiarities
(important for superqubit constructions). Indeed, the set of homomorphisms
$\left\{  \mathbf{T}\right\}  $ from superspace $\mathcal{V}$ to superspace
$\mathcal{W}$ is defined by the standard way%
\begin{equation}
\operatorname*{Hom}\left(  \mathcal{V},\mathcal{W}\right)  =\left\{
\mathbf{T}\mid\mathbf{T}\ \mathcal{V}\subset\mathcal{W}\right\}  .
\label{qs-ht}%
\end{equation}

We could assume that the $\mathbb{Z}_{2}$-graded structure is analogous to
(\ref{qs-vw})%
\begin{equation}
\operatorname*{Hom}\nolimits^{\left(  \mathsf{k}\right)  }\left(
\mathcal{V},\mathcal{W}\right)  =\left\{  \mathbf{T}\in\operatorname*{Hom}%
\left(  \mathcal{V},\mathcal{W}\right)  \mid\mathbf{T}\ \mathfrak{V}^{\left(
\mathsf{r}\right)  }\subset\mathfrak{W}^{\left(  \mathsf{r}\boxplus
\mathsf{k}\right)  }\right\}  ,\ \ \ \ \ \mathsf{k},\mathsf{r}\in
\mathbb{Z}_{2}. \label{qs-hk}%
\end{equation}

Note that only \textquotedblleft even\textquotedblright\ mappings
$\mathbf{T}^{\left(  \bar{0}\right)  }=\operatorname*{Hom}\nolimits^{\left(
\bar{0}\right)  }\left(  \mathcal{V},\mathcal{W}\right)  $ ($\deg
\mathbf{T}=\bar{0}\in\mathbb{Z}_{2}$) are homomorphisms. Only
\textquotedblleft odd\textquotedblright\ mappings $\mathbf{T}^{\left(  \bar
{1}\right)  }$ ($\deg\mathbf{T}=\bar{1}\in\mathbb{Z}_{2}$) are not morphisms
at all, because they cannot be composed: $\mathbf{T}^{\left(  \bar{1}\right)
}\circ\mathbf{T}^{\left(  \bar{1}\right)  }$ is not \textquotedblleft
odd\textquotedblright, but the \textquotedblleft even\textquotedblright\ mapping.

The same observation can be made for the linear operators in a $\mathbb{Z}%
_{2}$-graded linear vector space $\mathbb{C}^{p|q}$ given by (super)matrices.
In the standard basis a linear operator $\mathbf{T}\in\operatorname*{End}%
\left(  \mathbb{C}^{p|q}\right)  $ can be represented by the square block
$\left(  p+q\right)  \times\left(  p+q\right)  $ supermatrix over $\Lambda
_{N}\left(  \mathbb{C}\right)  $ \cite{berezin,lei1} (other representations
are also possible \cite{leites13})%
\begin{equation}
\mathrm{M}=\left(
\begin{array}
[c]{cc}%
A_{p\times p} & B_{p\times q}\\
C_{q\times p} & D_{q\times q}%
\end{array}
\right)  \in\mathrm{Mat}\left(  p|q,\Lambda_{N}\left(  \mathbb{C}\right)
\right)  , \label{qs-m}%
\end{equation}
where the even (ordinary) matrices $A_{p\times p}$, $D_{q\times q}$ are over
$\Lambda_{N}^{\left(  even\right)  }\left(  \mathbb{C}\right)  =\Lambda
_{N}^{\left(  \bar{0}\right)  }\left(  \mathbb{C}\right)  $, and the odd
(ordinary) matrices $B_{p\times q}$, $C_{q\times p}$ are over $\Lambda
_{N}^{\left(  odd\right)  }\left(  \mathbb{C}\right)  =\Lambda_{N}^{\left(
\bar{1}\right)  }\left(  \mathbb{C}\right)  $. Such full supermatrix has the
total parity (degree) $\deg\mathrm{M}=0$ and%
\begin{equation}
\mathrm{M}_{\deg\mathrm{M}=0}=\mathrm{M}_{body}+\mathrm{M}_{soul}=\left(
\begin{array}
[c]{cc}%
\left(  A_{p\times p}\right)  _{body} & 0_{p\times q}\\
0_{q\times p} & \left(  D_{q\times q}\right)  _{body}%
\end{array}
\right)  +\mathrm{M}_{soul}. \label{qs-md0}%
\end{equation}
If oppositely, $A_{p\times p}$, $D_{q\times q}$ are (ordinary) matrices over
$\Lambda_{N}^{\left(  \bar{1}\right)  }\left(  \mathbb{C}\right)  $, and
$B_{p\times q}$, $C_{q\times p}$ are (ordinary) matrices over $\Lambda
_{N}^{\left(  \bar{0}\right)  }\left(  \mathbb{C}\right)  $, then
$\deg\mathrm{M}=1$, and%
\begin{equation}
\mathrm{M}_{\deg\mathrm{M}=1}=\mathrm{M}_{body}+\mathrm{M}_{soul}=\left(
\begin{array}
[c]{cc}%
0_{p\times p} & \left(  B_{p\times q}\right)  _{body}\\
\left(  C_{q\times p}\right)  _{body} & 0_{q\times q}%
\end{array}
\right)  +\mathrm{M}_{soul}. \label{qs-md1}%
\end{equation}

Therefore, supermatrices with $\deg\mathrm{M}=1$ (only) are not morphisms of
$\mathbb{C}^{p|q}$, because their product gives the supermatrices having
$\deg\mathrm{M}=0$, and therefore%
\begin{equation}
\left\{  \mathrm{M}\right\}  _{\deg\mathrm{M}=1}\notin\operatorname*{End}%
\left(  \mathbb{C}^{p|q}\right)  .
\end{equation}

Thus, both supermatrices with $\deg\mathrm{M}=0$ and $\deg\mathrm{M}=1$
(considered together) are morphisms%
\begin{equation}
\left\{  \mathrm{M}\right\}  _{\deg\mathrm{M}=0}\cup\left\{  \mathrm{M}%
\right\}  _{\deg\mathrm{M}=1}\in\operatorname*{End}\left(  \mathbb{C}%
^{p|q}\right)  .\label{qs-mme}%
\end{equation}

After the decomposition of the matrices (\ref{qs-m}) with $\deg\mathrm{M}=0$
(reminding (\ref{qs-md0}), (\ref{qs-md1}))%
\begin{align}
\mathrm{M}  &  =\mathrm{M}^{\left(  even\right)  }+\mathrm{M}^{\left(
odd\right)  }=\mathrm{M}^{\left(  0\right)  }+\mathrm{M}^{\left(  1\right)
},\label{qs-mm}\\
\mathrm{M}^{\left(  0\right)  }  &  =\left(
\begin{array}
[c]{cc}%
A_{p\times p} & 0_{p\times q}\\
0_{q\times p} & D_{q\times q}%
\end{array}
\right)  ,\label{qs-m0}\\
\mathrm{M}^{\left(  1\right)  }  &  =\left(
\begin{array}
[c]{cc}%
0_{p\times p} & B_{p\times q}\\
C_{q\times p} & 0_{q\times q}%
\end{array}
\right)  , \label{qs-m1}%
\end{align}
we observe that $\mathrm{M}^{\left(  \bar{0}\right)  }\mathrm{M}%
^{\prime\left(  \bar{0}\right)  }=\mathrm{M}^{\prime\prime\left(  \bar
{0}\right)  }$, and therefore the corresponding operators are
\textquotedblleft even\textquotedblright\ endomorphisms of $\mathbb{C}^{p|q}$%
\begin{equation}
\left\{  \mathbf{T}^{\left(  \bar{0}\right)  }\right\}  \in\operatorname*{End}%
\left(  \mathbb{C}^{p|q}\right)  , \label{qs-te0}%
\end{equation}
but
\begin{equation}
\mathrm{M}^{\left(  \bar{1}\right)  }\mathrm{M}^{\prime\left(  \bar{1}\right)
}=\mathrm{M}^{\prime\prime\left(  \bar{0}\right)  }.
\end{equation}

The set $\left\{  \mathrm{M}^{\left(  \bar{1}\right)  }\right\}  $ is not
closed under composition (matrix multiplication), therefore the corresponding
(only) \textquotedblleft odd\textquotedblright\ operators $\mathbf{T}^{\left(
\bar{1}\right)  }$ are not morphisms by definition at all%
\begin{equation}
\left\{  \mathbf{T}^{\left(  \bar{1}\right)  }\right\}  \notin
\operatorname*{End}\left(  \mathbb{C}^{p|q}\right)  . \label{qs-te1}%
\end{equation}

By analogy with (\ref{qs-mme}) we have the statement that both
\textquotedblleft even\textquotedblright\ and \textquotedblleft
odd\textquotedblright\ superoperators are morphisms, such that%
\begin{equation}
\left\{  \mathbf{T}^{\left(  \bar{0}\right)  }\right\}  \cup\left\{
\mathbf{T}^{\left(  \bar{1}\right)  }\right\}  \in\operatorname*{End}\left(
\mathbb{C}^{p|q}\right)  .
\end{equation}

The above considerations should be taken into account during consistent
calculations with superqubits and supersymmetric quantum gates.

We remind some common notions in present notations for self-consistency.
First, as opposite to the standard transpose operator $\mathsf{T}%
:\mathrm{Mat}\left(  p,\mathbb{C}\right)  \rightarrow\mathrm{Mat}\left(
p,\mathbb{C}\right)  $, the supertranspose operator $\mathsf{sT}%
:\mathrm{Mat}\left(  p|q,\Lambda_{N}\left(  \mathbb{C}\right)  \right)
\rightarrow\mathrm{Mat}\left(  p|q,\Lambda_{N}\left(  \mathbb{C}\right)
\right)  $ is double-valued depending of the parity of supermatrix%
\begin{equation}
\left(
\begin{array}
[c]{cc}%
A_{p\times p} & B_{p\times q}\\
C_{q\times p} & D_{q\times q}%
\end{array}
\right)  ^{\mathsf{sT}}=\left\{
\begin{array}
[c]{c}%
\left(
\begin{array}
[c]{cc}%
A_{p\times p}^{\mathsf{T}} & C_{q\times p}^{\mathsf{T}}\\[5pt]%
-B_{p\times q}^{\mathsf{T}} & D_{q\times q}^{\mathsf{T}}%
\end{array}
\right)  ,\ \ \ if\ \ \deg M=\bar{0},\\[15pt]%
\left(
\begin{array}
[c]{cc}%
A_{p\times p}^{\mathsf{T}} & -C_{q\times p}^{\mathsf{T}}\\[5pt]%
B_{p\times q}^{\mathsf{T}} & D_{q\times q}^{\mathsf{T}}%
\end{array}
\right)  ,\ \ \ if\ \ \deg M=\bar{1}.
\end{array}
\right.  \label{qs-st}%
\end{equation}

It is seen that $\left(  \mathsf{sT}\right)  ^{\circ2}\neq\operatorname*{id}$
(while $\mathsf{T}^{\circ2}=\operatorname*{id}$), but $\left(  \mathsf{sT}%
\right)  ^{\circ4}=\operatorname*{id}$, and therefore the supertranspose
operator is the reflection of order $4$, while the transpose is the ordinary
reflection (of order $2$). For two supermatrices of the same shape
$\mathrm{M},\mathrm{N}\in\mathrm{Mat}\left(  p|q,\Lambda_{N}\left(
\mathbb{C}\right)  \right)  $ we have%
\begin{equation}
\left(  \mathrm{MN}\right)  ^{\mathsf{sT}}=\left(  -1\right)  ^{\deg
\mathrm{M}\deg\mathrm{N}}\mathrm{N}^{\mathsf{sT}}\mathrm{M}^{\mathsf{sT}},
\end{equation}
and in particular%
\begin{equation}
\left(  a\mathrm{M}\right)  ^{\mathsf{sT}}=a\mathrm{M}^{\mathsf{sT}%
},\ \ \ \ \forall a\in\Lambda_{N}\left(  \mathbb{C}\right)  ,
\end{equation}
which means that supertranspose $\mathsf{sT}$ is a $\Lambda_{N}\left(
\mathbb{C}\right)  $-module, as in case of the ordinary transpose operator
$\mathsf{T}$ which is a $\mathbb{C}$-module.

The supertrace is the homomorphism $\operatorname*{str}:\mathrm{Mat}\left(
p|q,\Lambda_{N}\left(  \mathbb{C}\right)  \right)  \rightarrow\Lambda
_{N}\left(  \mathbb{C}\right)  $ that is also double-valued (depending of
parity of supermatrix) mapping (for the supermatrix of the standard format
(\ref{qs-m}))%
\begin{equation}
\operatorname*{str}\left(
\begin{array}
[c]{cc}%
A_{p\times p} & B_{p\times q}\\
C_{q\times p} & D_{q\times q}%
\end{array}
\right)  =\left\{
\begin{array}
[c]{c}%
\operatorname*{tr}A_{p\times p}-\operatorname*{tr}D_{q\times q}%
,\ \ \ \text{if}\ \ \deg M=\bar{0},\\[3pt]%
\operatorname*{tr}A_{p\times p}+\operatorname*{tr}D_{q\times q}%
,\ \ \ \text{if}\ \ \deg M=\bar{1},
\end{array}
\right.
\end{equation}
which is additive and has the supercommutativity property, where $M,N$ are
ordinary matrices, and $\operatorname*{str}$ is invariant with repect to
supertranspose (analogous to ordinary trace)%
\begin{equation}%
\begin{tabular}
[c]{ccc}%
$\operatorname*{tr}M^{\mathsf{T}}=\operatorname*{tr}M$ & $\overset
{susy}{\Longrightarrow}$ & $\operatorname*{str}\mathrm{M}^{\mathsf{sT}%
}=\operatorname*{str}\mathrm{M}.$%
\end{tabular}
\end{equation}

The standard superdeterminant \cite{pakh,ber/lei} (or Berezinian
\cite{berezin,lei1}) in our notation is%
\begin{equation}
\operatorname*{Ber}\mathrm{M}=\operatorname*{Ber}\left(
\begin{array}
[c]{cc}%
A_{p\times p} & B_{p\times q}\\
C_{q\times p} & D_{q\times q}%
\end{array}
\right)  =\det\left(  A_{p\times p}-B_{p\times q}D_{q\times q}^{-1}C_{q\times
p}\right)  \left(  \det D_{q\times q}\right)  ^{-1}, \label{qs-ber}%
\end{equation}
which differs from the ordinary detrminant by the power $\left(  -1\right)  $
in the last multiplier. The mapping $\operatorname*{Ber}$ is a homomorphism of
$\mathrm{Mat}\left(  p|q,\Lambda_{N}\left(  \mathbb{C}\right)  \right)  $ and
invariant with respect to supertranspose $\mathsf{sT}$ (\ref{qs-st})%
\begin{equation}%
\begin{tabular}
[c]{ccc}%
$\det\left(  M^{\mathsf{T}}\right)  =\det M$ & $\overset{susy}{\Longrightarrow
}$ & $\operatorname*{Ber}\left(  \mathrm{M}^{\mathsf{sT}}\right)
=\operatorname*{Ber}\mathrm{M}.$%
\end{tabular}
\ \label{qs-dmt}%
\end{equation}

The connection of $\operatorname*{Ber}$ and $\operatorname*{str}$ is similar
to the ordinary case%
\begin{align}
&
\begin{tabular}
[c]{ccc}%
$\det M=e^{\operatorname*{tr}\left(  \ln M\right)  }$ & $\overset
{susy}{\Longrightarrow}$ & $\operatorname*{Ber}\mathrm{M}%
=e^{\operatorname*{str}\left(  \ln\mathrm{M}\right)  },$%
\end{tabular}
\label{qs-dme1}\\
&
\begin{tabular}
[c]{ccc}%
$\det e^{M}=e^{\operatorname*{tr}M}$ & $\overset{susy}{\Longrightarrow}$ &
$\operatorname*{Ber}e^{\mathrm{M}}=e^{\operatorname*{str}\mathrm{M}},$%
\end{tabular}
\ \label{qs-dme2}%
\end{align}
where $M\in\mathrm{Mat}\left(  p,\mathbb{C}\right)  $ and $\mathrm{M}%
\in\mathrm{Mat}\left(  p|q,\Lambda_{N}\left(  \mathbb{C}\right)  \right)  $.

The Berezinian (\ref{qs-ber}) has the inconvenient property for cheracterizing
the entanglement: $\operatorname*{Ber}$ is not defined for noninvertible
$D_{q\times q}$. Therefore, in \cite{bor/dah/duf/rub} it was proposed to use
for entanglement measure another possible function which has many properties
of Berezinian (but not all of them), satisfies (\ref{qs-dmt}) and has the
ordinary $\det$ as the nonsupersymmetric limit (when odd variables vanish).
Because the notion $\operatorname*{sdet}$ is widely used for
$\operatorname*{Ber}$ \cite{berezin,lei1}, we denote this function
$\mathrm{\operatorname*{sdTr}}$ which can be defined by the following informal
analogy%
\begin{equation}%
\begin{tabular}
[c]{ccc}%
$\det M=\dfrac{1}{2}\operatorname*{tr}\left(  \left(  ME_{sl}\right)
^{\mathsf{T}}\left(  ME_{sl}\right)  \right)  $ & $\overset{susy}%
{\Longrightarrow}$ & $\operatorname*{sdTr}\mathrm{M}=\dfrac{1}{2}%
\operatorname*{str}\left(  \left(  \mathrm{ME}_{osp}\right)  ^{\mathsf{sT}%
}\left(  \mathrm{ME}_{osp}\right)  \right)  ,$%
\end{tabular}
\end{equation}
where $E_{sl}$ and $\mathrm{E}_{osp}$ are $SL\left(  2\right)  $ and
$OSp\left(  1\mid2\right)  $ invariant tensors \cite{leites13} (determining
the corresponding group and supergroup in the standard way $M^{\mathsf{T}%
}E_{sl}M=M$ and $\mathrm{M}^{\mathsf{sT}}\mathrm{E}_{osp}\mathrm{M}%
=\mathrm{M}$). The main property of $\mathrm{\operatorname*{sdTr}}$ is
vanishing on the direct product states, and therefore it can measure, whether
a quantum (two superqubit) state is unentangled or entangled (see below).

\section{\textsc{Super Hilbert spaces and operators}}

Let us denote vectors (quantum states) in the $r$-dimensional complex Hilbert
space $\mathfrak{H}_{r}$ by kets $\left\vert \psi\right\rangle $ and the inner
product by $\left\langle \_|\_\right\rangle :$ $\mathfrak{H}_{r}%
\times\mathfrak{H}_{r}\rightarrow\mathbb{C}$, which is a non-degenerate
Hermitean and positive form that is linear in the first argument and
antilinear (conjugate linear) in the second one. The bra $\left\langle
\varphi\right\vert $ is defined as an element of the dual space $\mathfrak{H}%
_{r}^{\dagger}$ (in the notation of \cite{bor/dah/duf/rub}, for a inner
product vector space $\mathfrak{V}$ the notation $\mathfrak{V}^{\ast}$ is also
used for its dual), which is the functional $\left\langle \varphi
|\_\right\rangle :\mathfrak{H}_{r}\rightarrow\mathbb{C}$, such that the action
on a ket is denoted by $\left\langle \varphi\right\vert \left(  \left\vert
\psi\right\rangle \right)  :=\left\langle \varphi|\psi\right\rangle $ and
coincides with the inner product after the identification of the Hilbert space
with its dual (Riesz representation theorem \cite{rudin}). Informally, one can
write the injection $\left(  \left\vert \psi\right\rangle \right)  ^{\dagger
}=\left\langle \psi\right\vert $, which in the matrix representation (and
finite dimensional) standardly coincides with Hermitean adjoint, when
$\left\vert \psi\right\rangle $ becomes a matrix-column, $\left\langle
\psi\right\vert $ is a matrix-row, and the inner product is a scalar product,
also the obvious property $\left\langle \varphi|\psi\right\rangle ^{\dagger
}=\left\langle \psi|\varphi\right\rangle $ holds valid.

In a similar way, we consider the $\left(  r|s\right)  $-dimensional super
Hilbert space $\mathcal{H}_{\left(  r|s\right)  }$ over $\Lambda_{N}\left(
\mathbb{C}\right)  $ as the $\mathbb{Z}_{2}$-graded space $\mathcal{H}%
_{\left(  r|s\right)  }=\mathcal{H}_{\left(  r|s\right)  }^{\left(  \bar
{0}\right)  }\oplus\mathcal{H}_{\left(  r|s\right)  }^{\left(  \bar{1}\right)
}$, such that the supersymmetric quantum states, if homogenous, carry
$\mathbb{Z}_{2}$-grading $\pi_{\psi}=\bar{0},\bar{1}\in\mathbb{Z}_{2}$, and
they are denoted by super-kets $\Vert\mathit{\psi}^{\left(  \pi_{\psi}\right)
}\rangle\in\mathcal{H}_{\left(  r|s\right)  }^{\left(  \pi_{\psi}\right)  }$
with $\pi_{\psi}=\deg\mathit{\psi}$, while the body of even super states are
the ordinary kets%
\begin{equation}
\Vert\mathit{\psi}^{\left(  \bar{0}\right)  }\rangle_{body}=\left\vert
\psi\right\rangle \in\mathfrak{H}_{r}.
\end{equation}
We denote the super inner product by $\left\langle \_\Vert\_\right\rangle
:\mathcal{H}_{\left(  r|s\right)  }\times\mathcal{H}_{\left(  r|s\right)
}\rightarrow\Lambda_{N}\left(  \mathbb{C}\right)  $ obeying the property%
\begin{equation}
\left\langle \_\Vert\_\right\rangle _{body}=\left\langle \_|\_\right\rangle
\in\mathbb{C}.
\end{equation}

The super dual Hilbert space $\mathcal{H}_{\left(  r|s\right)  }^{\ddag}$ is
defined as the space of the functionals $\left\langle \mathit{\varphi
}^{\left(  \pi_{\varphi}\right)  }\Vert\_\right\rangle :\mathcal{H}_{\left(
r|s\right)  }^{\left(  \pi_{\psi}\right)  }\rightarrow\Lambda_{N}\left(
\mathbb{C}\right)  $, and the super bra $\langle\mathit{\varphi}^{\left(
\pi_{\varphi}\right)  }\Vert$ with $\pi_{\varphi}=\deg\mathit{\varphi}%
\in\mathbb{Z}_{2}$ is given by the action%
\begin{equation}%
\begin{tabular}
[c]{ccc}%
$\left\langle \varphi\right\vert \left(  \left\vert \psi\right\rangle \right)
=\left\langle \varphi|\psi\right\rangle \in\mathbb{C}$ & $\overset
{susy}{\Longrightarrow}$ & $\langle\mathit{\varphi}^{\left(  \pi_{\varphi
}\right)  }\Vert\left(  \Vert\mathit{\psi}^{\left(  \pi_{\psi}\right)
}\rangle\right)  =\delta_{\pi_{\varphi}\pi_{\psi}}\langle\mathit{\varphi
}^{\left(  \pi_{\varphi}\right)  }\Vert\mathit{\psi}^{\left(  \pi_{\psi
}\right)  }\rangle\in\Lambda_{N}\left(  \mathbb{C}\right)  .$%
\end{tabular}
\ \ \label{qs-pp}%
\end{equation}

The presence of the delta-function $\delta_{\pi_{\varphi}\pi_{\psi}}$ in
(\ref{qs-pp}) means the commonly used agreement that the graded super vectors
of opposite parity are \textquotedblleft mutually orthogonal\textquotedblright%
\begin{equation}
\langle\mathit{\varphi}^{\left(  \pi_{\varphi}\right)  }\Vert\mathit{\psi
}^{\left(  \pi_{\psi}\right)  }\rangle=0,\ \text{if\ }\pi_{\varphi}\neq
\pi_{\psi}. \label{qs-pp1}%
\end{equation}
Therefore, below in similar expressions we will put%
\begin{equation}
\pi_{\varphi}=\pi_{\psi}=\pi=\bar{0},\bar{1}\in\mathbb{Z}_{2}. \label{qs-pff}%
\end{equation}
In this case, we have%
\begin{equation}
\langle\mathit{\varphi}^{\left(  \pi\right)  }\Vert\mathit{\psi}^{\left(
\pi\right)  }\rangle^{\sharp}=\langle\mathit{\psi}^{\left(  \pi\right)  }%
\Vert\mathit{\varphi}^{\left(  \pi\right)  }\rangle.
\end{equation}
Thus, informally, one can write%
\begin{equation}
\left(  \Vert\mathit{\psi}^{\left(  \pi\right)  }\rangle\right)  ^{\ddag
}=\langle\mathit{\psi}^{\left(  \pi\right)  }\Vert,
\end{equation}
which means that $\left(  \ddag\right)  $ does not change parity $\pi$, and it
is the reflection of order $4$, because%
\begin{align}
\left(  \Vert\mathit{\psi}^{\left(  \pi\right)  }\rangle\right)  ^{\ddag
\ddag}  &  =\left(  -1\right)  ^{\pi}\Vert\mathit{\psi}^{\left(  \pi\right)
}\rangle,\\
\left(  \Vert\mathit{\psi}^{\left(  \pi\right)  }\rangle\right)  ^{\ddag
\ddag\ddag\ddag}  &  =\Vert\mathit{\psi}^{\left(  \pi\right)  }\rangle.
\end{align}

If $\mathit{z}\in\Lambda_{N}\left(  \mathbb{C}\right)  $ has a fixed parity,
then its product with super ket and super bra behaves with respect to $\left(
\ddag\right)  $ differently
\begin{align}
\left(  \Vert\mathit{\psi}^{\left(  \pi\right)  }\rangle\mathit{z}\right)
^{\ddag}  &  =\left(  -1\right)  ^{\pi\deg\mathit{z}}\mathit{z}^{\sharp
}\langle\mathit{\psi}^{\left(  \pi\right)  }\Vert,\\
\left(  \mathit{z}\langle\mathit{\psi}^{\left(  \pi\right)  }\Vert\right)
^{\ddag}  &  =\left(  -1\right)  ^{\pi\left(  \deg\mathit{z}+1\right)  }%
\Vert\mathit{\psi}^{\left(  \pi\right)  }\rangle\mathit{z}^{\sharp},
\end{align}
where $\left(  \sharp\right)  $ is the graded involution or superstar
(\ref{qs-sstar}).

We can omit the $\delta$-function in (\ref{qs-pp}), and this will lead to a
new kind of Hilbert spaces which allow mixing of gradings, such that all above
formulas should be changed.

In the super Hilbert space $\mathcal{H}_{\left(  r|s\right)  }$ the
superadjoint $\left(  \ddag\right)  $ of the superoperator (\ref{qs-ht}) with
the standard graded structure (\ref{qs-hk}) is defined by%
\begin{equation}%
\begin{tabular}
[c]{ccc}%
$\left\langle \mathit{T}\ \varphi|\psi\right\rangle =\left\langle
\varphi|\mathit{T}^{\dagger}\ \psi\right\rangle $ & $\overset{susy}%
{\Longrightarrow}$ & $\langle\mathbf{T}^{\left(  \pi_{T}\right)
}\ \mathit{\varphi}^{\left(  \pi_{\varphi}\right)  }\Vert\mathit{\psi
}^{\left(  \pi_{\psi}\right)  }\rangle=\left(  -1\right)  ^{\pi_{\varphi}%
\pi_{T}}\langle\mathit{\varphi}^{\left(  \pi_{\varphi}\right)  }%
\Vert\mathbf{T}^{\left(  \pi_{T}\right)  \ddag}\ \mathit{\psi}^{\left(
\pi_{\psi}\right)  }\rangle,$%
\end{tabular}
\ \ \ \label{qs-tf}%
\end{equation}
where $\mathit{T}$ and $\mathit{T}^{\dagger}$ are an operator and its adjoint
in the Hilbert space $\mathfrak{H}_{r}$, $\left\vert \psi\right\rangle
,\left\vert \varphi\right\rangle \in\mathfrak{H}_{r}$, and $\Vert
\mathit{\varphi}^{\left(  \pi_{\varphi}\right)  }\rangle\in\mathcal{H}%
_{\left(  r|s\right)  }^{\left(  \pi\right)  }$, $\Vert\mathit{\psi}^{\left(
\pi_{\psi}\right)  }\rangle\in\mathcal{H}_{\left(  r|s\right)  }^{\left(
\pi_{\psi}\right)  }$, $\pi_{T}=\deg\mathbf{T}\in\mathbb{Z}_{2}$. Here we do
not have the restriction (\ref{qs-pff}), because the superoperator
$\mathbf{T}^{\left(  \pi_{T}\right)  }$ with $n_{T}=\bar{1}$ can change the
parity of quantum states. The superadjoint of the action on the quantum state
is%
\begin{equation}%
\begin{tabular}
[c]{ccc}%
$\left(  \mathit{T}\ \left\vert \psi\right\rangle \right)  ^{\dagger
}=\left\langle \psi\right\vert \ \mathit{T}^{\dagger}$ & $\overset
{susy}{\Longrightarrow}$ & $\left(  \mathbf{T}^{\left(  \pi_{T}\right)  }%
\Vert\mathit{\psi}^{\left(  \pi_{\psi}\right)  }\rangle\right)  ^{\ddag
}=\left(  -1\right)  ^{\pi_{\psi}\pi_{T}}\langle\mathit{\psi}^{\left(
\pi_{\psi}\right)  }\Vert\mathbf{T}^{\left(  \pi_{T}\right)  \ddag},$%
\end{tabular}
\end{equation}
The definition (\ref{qs-tf}) is equivalent to%
\begin{equation}
\langle\mathit{\varphi}^{\left(  \pi_{\varphi}\right)  }\Vert\mathbf{T}%
^{\left(  \pi_{T}\right)  \ddag}\Vert\mathit{\psi}^{\left(  \pi_{\psi}\right)
}\rangle=\left(  -1\right)  ^{\pi_{\varphi}\pi_{\psi}+\pi_{\psi}+\left(
\pi_{\varphi}+\pi_{\psi}\right)  \pi_{T}}\langle\mathit{\psi}^{\left(
\pi_{\psi}\right)  }\Vert\mathbf{T}^{\left(  \pi_{T}\right)  }\Vert
\mathit{\varphi}^{\left(  \pi_{\varphi}\right)  }\rangle^{\sharp}.
\end{equation}

If the superoperator $\mathbf{T}$ has a supermatrix representation in
$\mathrm{Mat}\left(  p|q,\Lambda_{N}\left(  \mathbb{C}\right)  \right)  $, the
its superadjoint is represented by composition of supertranspose (\ref{qs-st})
and the graded involution (superstar) (\ref{qs-sstar}) as%
\begin{equation}%
\begin{tabular}
[c]{ccc}%
$M^{\dagger}=\overline{M}^{\mathsf{T}},\ \ M\in\mathrm{Mat}\left(
p,\mathbb{C}\right)  $ & $\overset{susy}{\Longrightarrow}$ & $\mathrm{M}%
^{\ddag}=\left(  \mathrm{M}^{\sharp}\right)  ^{\mathsf{sT}},$ $\mathrm{M}%
\in\mathrm{Mat}\left(  p|q,\Lambda_{N}\left(  \mathbb{C}\right)  \right)  ,$%
\end{tabular}
\end{equation}
which is the superanalog of the Hermitean conjugation (conjugate transpose).

\section{\textsc{Qubits and superqubits}}

Mathematically qubits (or $d$-qudits) and superqubits (or $\left(  r|s\right)
$-superqudits) are normalized vectors in the $r$-dimensional Hilbert space and
$\left(  r|s\right)  $-dimensional super Hilbert space, respectively, which
are presented in the Dirac bra-ket notation (see previous Section). They are
written in the computational basis to study thoroughly various symmetries and
introduce suitable variables which can measure entanglement in consistent
ways. Because the super Hilbert space is $\mathbb{Z}_{2}$-graded, there can
exist even and odd vectors (as for the general quantum states in the previos
section) which can correspond to even and odd superqubits, respectively.

The definitions of a single qudit in the complex Hilbert space $\mathfrak{H}%
_{d}$ and a single superqudit in super Hilbert space $\mathcal{H}_{\left(
r|s\right)  }$ (over $\Lambda_{N}\left(  \mathbb{C}\right)  $) can be written,
in general, as the expansions of the (pure) quantum states on the
computational (super) basis as follows%
\begin{align}
&  \left\vert \Psi\right\rangle =\left\vert \Psi\right\rangle _{\left(
d\right)  }=x_{0}\left\vert 0\right\rangle +x_{1}\left\vert 1\right\rangle
+\ldots+x_{d-1}\left\vert d-1\right\rangle ,\label{qs-qu}\\
&  \left\vert x\right\vert _{0}^{2}+\left\vert x\right\vert _{1}^{2}%
+\ldots+\left\vert x\right\vert _{d-1}^{2}=1,\ \ \ \ x_{i}\in\mathbb{C}%
,\left\vert i\right\rangle \in\mathfrak{H}_{d},i=0,\ldots,d-1,\label{qs-qu1}\\
&  \Downarrow_{susy}\nonumber\\
&  \Vert\mathbf{\Psi}^{\left(  \bar{0}\right)  }\rangle=\Vert\mathbf{0\rangle
\mathit{x}_{0}+\Vert1\mathbf{\rangle\mathit{x}_{1}+}\ldots+}\Vert
\mathbf{r}-1\mathbf{\rangle}\mathit{x}_{r-1}+\Vert\text{\textbf{\oe }}%
_{0}\rangle\text{\ae }_{0}+\ldots+\Vert\text{\textbf{\oe }}_{s-1}%
\rangle\text{\ae }_{s-1},\label{qs-squ}\\
&  \mathit{x}_{0}^{\sharp}\mathit{x}_{0}+\mathit{x}_{1}^{\sharp}\mathit{x}%
_{1}\mathbf{\mathbf{+}\ldots+}\mathit{x}_{r-1}^{\sharp}\mathit{x}%
_{r-1}-\text{\ae }_{0}^{\sharp}\text{\ae }_{0}-\ldots\text{\ae }_{s-1}%
^{\sharp}\text{\ae }_{s-1}=1,\label{qs-squ1}\\
&  \Vert\mathbf{\Psi}^{\left(  \bar{1}\right)  }\rangle=\Vert\mathbf{0\rangle
}\text{\ae }_{0}\mathbf{+\Vert1\mathbf{\rangle}}\text{\ae }%
\mathbf{\mathbf{_{1}+}\ldots+}\Vert\mathbf{r}-1\mathbf{\rangle}\text{\ae }%
_{r-1}+\Vert\text{\textbf{\oe }}_{0}\rangle\mathbf{\mathit{x}_{0}}%
+\ldots+\Vert\text{\textbf{\oe }}_{s-1}\rangle\mathit{x}_{s-1},\label{qs-squ2}%
\\
&  \mathit{x}_{i}\in\Lambda_{N}^{\left(  0\right)  }\left(  \mathbb{C}\right)
,\ \Vert\mathbf{j}\rangle\in\mathcal{H}_{\left(  r|s\right)  }^{\left(
\bar{0}\right)  },\ \text{\ae }_{\alpha}\in\Lambda_{N}^{\left(  \bar
{1}\right)  }\left(  \mathbb{C}\right)  ,\ \Vert\text{\textbf{\oe }}_{\alpha
}\rangle\in\mathcal{H}_{\left(  r|s\right)  }^{\left(  \bar{1}\right)
}.\nonumber
\end{align}

We assume that $\deg\Vert\mathbf{i}\rangle=\deg\mathbf{\mathbf{\mathit{x}}%
}_{i}=\bar{0}$, $\deg\Vert$\textbf{\oe }$_{\alpha}\rangle=\deg$\ae $_{\alpha
}=\bar{1}$, and therefore the superqudit (\ref{qs-squ}) has the even parity
$\pi_{\Psi}=\deg\Vert\mathbf{\Psi}\rangle=\bar{0}$, and we call it the even
superqudit $\Vert\mathbf{\Psi}^{\left(  \bar{0}\right)  }\rangle
=\Vert\mathbf{\Psi}\rangle^{even}$, while the superqudit (\ref{qs-squ2}) has
the odd parity $\pi_{\Psi}=\deg\Vert\mathbf{\Psi}\rangle=\bar{1}$, and we call
it the odd superqudit $\Vert\mathbf{\Psi}^{\left(  \bar{1}\right)  }%
\rangle=\Vert\mathbf{\Psi}\rangle^{odd}$, denoting both of them $\Vert
\mathbf{\Psi}^{\left(  \pi\right)  }\rangle=\Vert\mathbf{\Psi}^{\left(
\pi_{\Psi}\right)  }\rangle$. The normalization of the odd superqudit can be
done using some special Grassmann norms considered in
\cite{rud2000,rogers,hab/kup}.

\begin{definition}
\label{qs-def-span}The qudits $\left\vert \Psi\right\rangle $ and superqudits
$\Vert\mathbf{\Psi}\rangle$ are

\begin{enumerate}
\item linear spans of the corresponding subspace $\operatorname*{span}\left(
\left\{  \left\vert i\right\rangle \right\}  \right)  \subseteq\mathfrak{H}%
_{d}$ and subsuperspace $\operatorname*{span}\left(  \left\{  \Vert
\mathbf{i}\rangle\right\}  |\left\{  \Vert\text{\textbf{\oe }}_{\alpha}%
\rangle\right\}  \right)  \subseteq\mathcal{H}_{\left(  r|s\right)  }$, respectively,

\item having the normalization conditions (\ref{qs-qu1}), (\ref{qs-squ1}).
\end{enumerate}
\end{definition}

For consistency, it is natural to assume that the superqudit (\ref{qs-squ})
has the Grassmannless limit (body map \cite{rog1}) as the ordinary qudit
(\ref{qs-qu})%
\begin{equation}
\Vert\mathbf{\Psi}^{\left(  \bar{0}\right)  }\rangle_{body}=\left\vert
\Psi\right\rangle _{\left(  r\right)  }. \label{qs-red}%
\end{equation}

The normalization conditions (\ref{qs-qu1}), (\ref{qs-squ1}) distinguish
(super)qudits among general span subspaces, which allows us to endow them
probablistic interpretation. If the limit (\ref{qs-red}) is accepted, then
(\ref{qs-qu1}) and (\ref{qs-squ1}), as well as the basises $\left\{
\left\vert i\right\rangle \right\}  \in\mathfrak{H}_{r}$ and $\left\{
\Vert\mathbf{i}\rangle\right\}  \in\mathcal{H}_{\left(  r|s\right)  }$ are
connected with the body map.

The (super)qudits in minimum dimensions $d=2$, $r=2$ , $s=1$ are called
(super)qubits \cite{bor/dah/duf/rub} and have the form\footnote{For clarity
and convenience for applications we use the manifest presentation of different
variables. The right coordinates are used in superqubits according to the sign
agreement of \cite{bor/dah/duf/rub}.}%

\begin{equation}%
\begin{tabular}
[c]{ccc}%
$\left.
\begin{array}
[c]{c}%
\left\vert \Psi\right\rangle =\left\vert \Psi\right\rangle _{\left(  2\right)
}=x_{0}\left\vert 0\right\rangle +x_{1}\left\vert 1\right\rangle ,\\
\left\vert x_{0}\right\vert ^{2}+\left\vert x_{1}\right\vert ^{2}=1,\\
x_{0},x_{1}\in\mathbb{C},\ \ \ \ \left\vert 0\right\rangle ,\left\vert
1\right\rangle \in\mathfrak{H}_{2},
\end{array}
\right.  $ & $\overset{susy}{\Longrightarrow}$ & $\left.
\begin{array}
[c]{c}%
\Vert\mathbf{\Psi}^{\left(  \bar{0}\right)  }\rangle=\Vert\mathbf{0\rangle
\mathit{x}_{0}+\Vert1\mathbf{\rangle\mathit{x}_{1}+}}\Vert\text{\textbf{\oe }%
}\rangle\text{\ae },\\
\mathit{x}_{0}^{\sharp}\mathit{x}_{0}+\mathit{x}_{1}^{\sharp}\mathit{x}%
_{1}-\text{\ae }^{\sharp}\text{\ae }=1,\\
\Vert\mathbf{\Psi}^{\left(  \bar{1}\right)  }\rangle=\Vert\mathbf{0\rangle
}\text{\ae }_{0}\mathbf{+\Vert1\mathbf{\rangle}}\text{\ae }_{1}%
\mathbf{\mathbf{+}}\Vert\text{\textbf{\oe }}\rangle\mathit{x},\\
\mathit{x},\mathit{x}_{0},\mathit{x}_{1}\in\Lambda_{N}^{\left(  \bar
{0}\right)  }\left(  \mathbb{C}\right)  ,\ \ \ \ \Vert\mathbf{0}\rangle
,\Vert\mathbf{1}\rangle\in\mathcal{H}_{\left(  2|1\right)  }^{\left(  \bar
{0}\right)  },\\
\text{\ae },\text{\ae }_{0},\text{\ae }_{1}\in\Lambda_{N}^{\left(  \bar
{1}\right)  }\left(  \mathbb{C}\right)  ,\ \ \ \ \ \ \Vert\text{\textbf{\oe }%
}\rangle\in\mathcal{H}_{\left(  2|1\right)  }^{\left(  \bar{1}\right)  }.
\end{array}
\right.  $%
\end{tabular}
\ \ \label{qs-qd1}%
\end{equation}

There are four main operations between two single (super)qubits.

\begin{enumerate}
\item \textit{Inner product} of bra (super)qubit and ket (super)qubit%
\begin{equation}%
\begin{tabular}
[c]{ccc}%
$\left\langle \Psi|\Psi^{\prime}\right\rangle =\bar{x}_{0}x_{0}^{\prime}%
+\bar{x}_{1}x_{1}^{\prime}\in\mathbb{C}$ & $\overset{susy}{\Longrightarrow}$ &
$\left.
\begin{array}
[c]{c}%
\langle\mathbf{\Psi}^{\left(  \bar{0}\right)  }\Vert\mathbf{\Psi}^{\left(
\bar{0}\right)  \prime}\rangle=\mathit{x}_{0}^{\sharp}\mathit{x}_{0}^{\prime
}+\mathit{x}_{1}^{\sharp}\mathit{x}_{1}^{\prime}-\text{\ae }^{\sharp
}\text{\ae }^{\prime},\\
\langle\mathbf{\Psi}^{\left(  \bar{1}\right)  }\Vert\mathbf{\Psi}^{\left(
\bar{1}\right)  \prime}\rangle=\text{\ae }_{0}^{\sharp}\text{\ae }_{0}%
^{\prime}+\text{\ae }_{1}^{\sharp}\text{\ae }_{1}^{\prime}+\mathit{x}^{\sharp
}\mathit{x}^{\prime}%
\end{array}
\right.  $%
\end{tabular}
\label{qs-pi}%
\end{equation}
where $\overline{\left(  {}\right)  }$ is complex conjugation and $\left(
\sharp\right)  $ is the grade involution (\ref{qs-sstar}).

If the states coincide, $\left\vert \Psi^{\prime}\right\rangle =\left\vert
\Psi\right\rangle $ and $\Vert\mathbf{\Psi}^{\prime}\rangle=\Vert\mathbf{\Psi
}\rangle$, then (\ref{qs-pi}) are square norms of $\left\vert \Psi
\right\rangle $ and $\Vert\mathbf{\Psi}^{\left(  \bar{0}\right)  }\rangle$
becoming unity for normalized (super)qubits. For physical states the square
norm of the even superqubit is positive%
\begin{equation}
\Vert\mathbf{\Psi}^{\left(  \bar{0}\right)  }\mathbf{\Vert}_{body}^{2}%
=\langle\mathbf{\Psi}^{\left(  \bar{0}\right)  }\Vert\mathbf{\Psi}^{\left(
\bar{0}\right)  \prime}\rangle_{body}>0.
\end{equation}

\item \textit{Outer product} of ket and bra gives the density (super)matrix of
(super)qubit%
\begin{align}
\rho &  =\left\vert \Psi\right\rangle \left\langle \Psi\right\vert =\left(
\begin{array}
[c]{cc}%
x_{0}\bar{x}_{0} & x_{1}\bar{x}_{0}\\
x_{0}\bar{x}_{1} & x_{1}\bar{x}_{1}%
\end{array}
\right) \\
&  \Downarrow_{susy}\\
\mathbf{\rho}^{\left(  \bar{0}\right)  }  &  =\Vert\mathbf{\Psi}^{\left(
\bar{0}\right)  }\rangle\langle\mathbf{\Psi}^{\left(  \bar{0}\right)  }%
\Vert=\left(
\begin{array}
[c]{ccc}%
\mathit{x}_{0}\mathit{x}_{0}^{\sharp} & \mathit{x}_{1}\mathit{x}_{0}^{\sharp}
& \text{\ae }\mathit{x}_{0}^{\sharp}\\
\mathit{x}_{0}\mathit{x}_{1}^{\sharp} & \mathit{x}_{1}\mathit{x}_{1}^{\sharp}
& \text{\ae }\mathit{x}_{1}^{\sharp}\\
-\mathit{x}_{0}\text{\ae }^{\sharp} & -\mathit{x}_{1}\text{\ae }^{\sharp} &
-\text{\ae \ae }^{\sharp}%
\end{array}
\right)  ,\\
\mathbf{\rho}^{\left(  \bar{1}\right)  }  &  =\Vert\mathbf{\Psi}^{\left(
\bar{1}\right)  }\rangle\langle\mathbf{\Psi}^{\left(  \bar{1}\right)  }%
\Vert=\left(
\begin{array}
[c]{ccc}%
-\text{\ae }_{0}\text{\ae }_{0}^{\sharp} & -\text{\ae }_{1}\text{\ae }%
_{0}^{\sharp} & -\mathit{x}\text{\ae }_{0}^{\sharp}\\
-\text{\ae }_{0}\text{\ae }_{1}^{\sharp} & -\text{\ae }_{1}\text{\ae }%
_{1}^{\sharp} & -\mathit{x}\text{\ae }_{1}^{\sharp}\\
\text{\ae }_{0}\mathit{x}^{\sharp} & \text{\ae }_{1}\mathit{x}^{\sharp} &
\mathit{xx}^{\sharp}%
\end{array}
\right)  ,
\end{align}
and the body map limit for $\mathbf{\rho}^{\left(  \bar{0}\right)  }$ is
similar to (\ref{qs-red}). The standard connection of the inner product with
the (super)trace of density matrix for a given (super)qubit holds valid
(taking into account gradings)%
\begin{equation}%
\begin{tabular}
[c]{ccc}%
$\operatorname*{tr}\rho=\left\langle \Psi|\Psi\right\rangle \in\mathbb{C}$ &
$\overset{susy}{\Longrightarrow}$ & $\left.
\begin{array}
[c]{c}%
\operatorname*{str}\mathbf{\rho}^{\left(  \bar{0}\right)  }=\langle
\mathbf{\Psi}^{\left(  \bar{0}\right)  }\Vert\mathbf{\Psi}^{\left(  \bar
{0}\right)  }\rangle\in\Lambda_{N}^{\left(  \bar{0}\right)  }\left(
\mathbb{C}\right)  ,\\
\operatorname*{str}\mathbf{\rho}^{\left(  \bar{1}\right)  }=-\langle
\mathbf{\Psi}^{\left(  \bar{1}\right)  }\Vert\mathbf{\Psi}^{\left(  \bar
{1}\right)  }\rangle\in\Lambda_{N}^{\left(  \bar{0}\right)  }\left(
\mathbb{C}\right)  .
\end{array}
\right.  $%
\end{tabular}
\end{equation}

\item \textit{Tensor product} of two ket (super)qubits (or two bra
(super)qubits) (\ref{qs-qd1}) can be presented as the following manifest
expansions on elementary tensors (all gradings appear and are shown for
clarity and direct usage in computations)%
\begin{align}
&  \left\vert \Psi\right\rangle \otimes\left\vert \Psi^{\prime}\right\rangle
=x_{0}x_{0}^{\prime}\left\vert 0\right\rangle \otimes\left\vert 0^{\prime
}\right\rangle +x_{0}x_{1}^{\prime}\left\vert 0\right\rangle \otimes\left\vert
1^{\prime}\right\rangle +x_{1}x_{0}^{\prime}\left\vert 1\right\rangle
\otimes\left\vert 0^{\prime}\right\rangle +x_{1}x_{1}^{\prime}\left\vert
1\right\rangle \otimes\left\vert 1^{\prime}\right\rangle \nonumber\\
&  \Downarrow_{susy}\nonumber\\
&  \left(
\begin{array}
[c]{c}%
\Vert\mathbf{\Psi}^{\left(  \bar{0}\right)  }\rangle\otimes\Vert\mathbf{\Psi
}^{\left(  \bar{0}\right)  \prime}\rangle\\
\Vert\mathbf{\Psi}^{\left(  \bar{1}\right)  }\rangle\otimes\Vert\mathbf{\Psi
}^{\left(  \bar{1}\right)  \prime}\rangle\\
\Vert\mathbf{\Psi}^{\left(  \bar{0}\right)  }\rangle\otimes\Vert\mathbf{\Psi
}^{\left(  \bar{1}\right)  \prime}\rangle\\
\Vert\mathbf{\Psi}^{\left(  \bar{1}\right)  }\rangle\otimes\Vert\mathbf{\Psi
}^{\left(  \bar{0}\right)  \prime}\rangle
\end{array}
\right)  =\Vert\mathbf{0\rangle\otimes}\Vert\mathbf{0}^{\prime}\mathbf{\rangle
\left(
\begin{array}
[c]{c}%
\mathit{x}_{0}\mathit{x}_{0}^{\prime}\\
\text{\ae }_{0}\text{\ae }_{0}^{\prime}\\
\mathit{x}_{0}\text{\ae }_{0}^{\prime}\\
\text{\ae }_{0}\mathit{x}_{0}^{\prime}%
\end{array}
\right)  +}\Vert\mathbf{0\rangle\otimes}\Vert\mathbf{1}^{\prime}%
\mathbf{\rangle}\left(
\begin{array}
[c]{c}%
\mathit{x}_{0}\mathit{x}_{1}^{\prime}\\
\text{\ae }_{0}\text{\ae }_{1}^{\prime}\\
\mathit{x}_{0}\text{\ae }_{1}^{\prime}\\
\text{\ae }_{0}\mathit{x}_{1}^{\prime}%
\end{array}
\right) \nonumber\\
&  \mathbf{\mathbf{+}\Vert\mathbf{0\rangle\otimes}\Vert}\text{\textbf{\oe }%
}\mathbf{^{\prime}\mathbf{\rangle\left(
\begin{array}
[c]{c}%
\mathit{x}_{0}\text{\ae }^{\prime}\\
\text{\ae }_{0}\mathit{x}^{\prime}\\
\mathit{x}_{0}\mathit{x}^{\prime}\\
\text{\ae }_{0}\text{\ae }^{\prime}%
\end{array}
\right)  +}}\Vert\mathbf{1\rangle\otimes}\Vert\mathbf{0}^{\prime
}\mathbf{\rangle\left(
\begin{array}
[c]{c}%
\mathit{x}_{1}\mathit{x}_{0}^{\prime}\\
\text{\ae }_{1}\text{\ae }_{0}^{\prime}\\
\mathit{x}_{1}\text{\ae }_{0}^{\prime}\\
\text{\ae }_{1}\mathit{x}_{0}^{\prime}%
\end{array}
\right)  +}\Vert\mathbf{1\rangle\otimes}\Vert\mathbf{1}^{\prime}%
\mathbf{\rangle}\left(
\begin{array}
[c]{c}%
\mathit{x}_{1}\mathit{x}_{1}^{\prime}\\
\text{\ae }_{1}\text{\ae }_{1}^{\prime}\\
\mathit{x}_{1}\text{\ae }_{1}^{\prime}\\
\text{\ae }_{1}\mathit{x}_{1}^{\prime}%
\end{array}
\right) \nonumber\\
&  \mathbf{\mathbf{+}\Vert1\mathbf{\rangle\otimes}\Vert}\text{\textbf{\oe }%
}\mathbf{^{\prime}\mathbf{\rangle\left(
\begin{array}
[c]{c}%
\mathit{x}_{1}\text{\ae }^{\prime}\\
\text{\ae }_{1}\mathit{x}^{\prime}\\
\mathit{x}_{1}\mathit{x}^{\prime}\\
\text{\ae }_{1}\text{\ae }^{\prime}%
\end{array}
\right)  +}}\Vert\text{\textbf{\oe }}\mathbf{\rangle\otimes}\Vert
\mathbf{0}^{\prime}\mathbf{\rangle\left(
\begin{array}
[c]{c}%
\text{\ae }\mathit{x}_{0}^{\prime}\\
\mathit{x}\text{\ae }_{0}^{\prime}\\
\text{\ae \ae }_{0}^{\prime}\\
\mathit{xx}_{0}^{\prime}%
\end{array}
\right)  +}\Vert\text{\textbf{\oe }}\mathbf{\rangle\otimes}\Vert
\mathbf{1}^{\prime}\mathbf{\rangle}\left(
\begin{array}
[c]{c}%
\text{\ae }\mathit{x}_{1}^{\prime}\\
\mathit{x}\text{\ae }_{1}^{\prime}\\
\text{\ae \ae }_{1}^{\prime}\\
\mathit{xx}_{1}^{\prime}%
\end{array}
\right) \nonumber\\
&  \mathbf{+}\Vert\text{\textbf{\oe }}\mathbf{\rangle\otimes}\Vert
\text{\textbf{\oe }}^{\prime}\mathbf{\rangle}\left(
\begin{array}
[c]{c}%
\text{\ae \ae }^{\prime}\\
\mathit{xx}^{\prime}\\
\text{\ae }\mathit{x}^{\prime}\\
\mathit{x}\text{\ae }^{\prime}%
\end{array}
\right)  \mathbf{\in}\left(
\begin{array}
[c]{c}%
\left(  \mathcal{H}_{\left(  2|1\right)  }\otimes\mathcal{H}_{\left(
2|1\right)  }\right)  ^{\left(  \bar{0}\right)  }\\
\left(  \mathcal{H}_{\left(  2|1\right)  }\otimes\mathcal{H}_{\left(
2|1\right)  }\right)  ^{\left(  \bar{0}\right)  }\\
\left(  \mathcal{H}_{\left(  2|1\right)  }\otimes\mathcal{H}_{\left(
2|1\right)  }\right)  ^{\left(  \bar{1}\right)  }\\
\left(  \mathcal{H}_{\left(  2|1\right)  }\otimes\mathcal{H}_{\left(
2|1\right)  }\right)  ^{\left(  \bar{1}\right)  }%
\end{array}
\right)  . \label{qs-tp}%
\end{align}

Thus, there are 4 different superqubit tensor products depending of their parity.

\begin{definition}
The (pure) quantum state which can be obtained as a tensor product is called a
separable state.
\end{definition}

\item \textit{Cross product} of two ket qutrits ((\ref{qs-qu}) with $d=3$) of
the form%
\begin{align}
&  \left\vert \Psi\right\rangle =\left\vert \Psi\right\rangle _{\left(
3\right)  }=x_{0}\left\vert 0\right\rangle +x_{1}\left\vert 1\right\rangle
+x_{2}\left\vert 2\right\rangle ,\label{qs-qtr}\\
&  \left\vert x_{0}\right\vert ^{2}+\left\vert x_{1}\right\vert ^{2}%
+\left\vert x_{2}\right\vert ^{2}=1,\label{qs-qtr1}\\
&  x_{0},x_{1},x_{2}\in\mathbb{C},\ \ \ \ \left\vert 0\right\rangle
,\left\vert 1\right\rangle ,\left\vert 2\right\rangle \in\mathfrak{H}%
_{3},\nonumber
\end{align}
can be defined by analogy with ordinary cross product of vectors%
\begin{align}
&  \left\vert \Phi\right\rangle _{cross}=\left\vert \Psi\right\rangle
\times\left\vert \Psi^{\prime}\right\rangle =\sum_{i,j,k=0,1,2}\epsilon
_{ijk}x_{j}x_{k}^{\prime}\left\vert i\right\rangle \label{qs-cr0}\\
&  =\det\left(
\begin{array}
[c]{ccc}%
\left\vert 0\right\rangle  & \left\vert 1\right\rangle  & \left\vert
2\right\rangle \\
x_{0} & x_{1} & x_{2}\\
x_{0}^{\prime} & x_{1}^{\prime} & x_{2}^{\prime}%
\end{array}
\right)  =\det M_{\left\vert 0\right\rangle }\left\vert 0\right\rangle +\det
M_{\left\vert 1\right\rangle }\left\vert 1\right\rangle +\det M_{\left\vert
2\right\rangle }\left\vert 2\right\rangle \label{qs-fn}\\
&  =\left(  x_{1}x_{2}^{\prime}-x_{2}x_{1}^{\prime}\right)  \left\vert
0\right\rangle -\left(  x_{0}x_{2}^{\prime}-x_{2}x_{0}^{\prime}\right)
\left\vert 1\right\rangle +\left(  x_{0}x_{1}^{\prime}-x_{1}x_{0}^{\prime
}\right)  \left\vert 2\right\rangle ,\ \ x_{i},x_{i}^{\prime}\in\mathbb{C},
\label{qs-cr}%
\end{align}
where $M_{\left\vert i\right\rangle }$ is the minor of element $\left\vert
i\right\rangle $, and $\epsilon_{ijk}$ fully antisymmetric tensor. The last
expanded form (\ref{qs-cr}) is convenient to use for superqubits as well.

\begin{definition}
We call the qutrit $\left\vert \Phi\right\rangle _{cross}$ which is built as
the cross product (\ref{qs-cr}) a cross-qutrit.
\end{definition}

The square norm of the cross-qutrit is%
\begin{align}
\left\Vert \left\vert \Phi\right\rangle _{cross}\right\Vert ^{2}  &
=\left\Vert \Psi\right\Vert ^{2}\left\Vert \Psi^{\prime}\right\Vert
^{2}-\left\vert \left\langle \Psi|\Psi^{\prime}\right\rangle \right\vert
^{2}\nonumber\\
&  =\left\vert \det M_{\left\vert 0\right\rangle }\right\vert ^{2}+\left\vert
\det M_{\left\vert 1\right\rangle }\right\vert ^{2}+\left\vert \det
M_{\left\vert 2\right\rangle }\right\vert ^{2}. \label{qs-fnn}%
\end{align}
Therefore, for the normalized qutrit $\left\vert \Phi\right\rangle _{cross}$
we have the additional condition (together with 2 ones (\ref{qs-qtr1}) for
$\left\vert \Psi\right\rangle $ and $\left\vert \Psi^{\prime}\right\rangle $)%
\begin{equation}
\left\vert \det M_{\left\vert 0\right\rangle }\right\vert ^{2}+\left\vert \det
M_{\left\vert 1\right\rangle }\right\vert ^{2}+\left\vert \det M_{\left\vert
2\right\rangle }\right\vert ^{2}=1.
\end{equation}

\begin{definition}
The (pure) quantum state which can be obtained as a cross product is called a
cross-separable state.
\end{definition}

The cross-qutrits have special properties and can be connected with the
concurrence measure in considering entanglement (see below).\bigskip
\end{enumerate}

\section{\textsc{Multi-(super)qubit states}}

The multi-(super)qudit quantum states are vectors in the tensor product of $n$
(super) Hilbert spaces $\mathfrak{H}_{d}^{\otimes n}=\overset{n}%
{\overbrace{\mathfrak{H}_{d}\otimes\ldots\otimes\mathfrak{H}_{d}}}$ (resp.
$\mathcal{H}_{\left(  r|s\right)  }^{\otimes n}=\overset{n}{\overbrace
{\mathcal{H}_{\left(  r|s\right)  }\otimes\ldots\otimes\mathcal{H}_{\left(
r|s\right)  }}}$). From the first sight, a straightforward way to obtain such
vectors is to use the tensor product (\ref{qs-tp}) repeatedly $n-1$ times.
However, such procedure is too restricted and can give separable states only.
The consequent definition should be made in terms of $\operatorname*{span}$'s
as \textbf{Definition} \ref{qs-def-span}.

\begin{definition}
The multi-qudits ($n$-qudit states) are

\begin{enumerate}
\item linear span of the Hilbert subspace%
\begin{align}
\left\{  \left\vert \Psi\left(  n\right)  \right\rangle \right\}   &
=\operatorname*{span}(\left\vert i_{1}\right\rangle \otimes\ldots
\otimes\left\vert i_{n}\right\rangle )\subseteq\overset{n}{\overbrace
{\mathfrak{H}_{d}\otimes\ldots\otimes\mathfrak{H}_{d}}},\ \ \ \left\vert
i_{k}\right\rangle \in\mathfrak{H}_{d},\ \ \ k=1,\ldots,n,\nonumber\\
\left\vert \Psi\left(  n\right)  \right\rangle  &  =%
{\displaystyle\sum\limits_{i_{1}=0}^{d-1}}
\ldots%
{\displaystyle\sum\limits_{i_{n}=0}^{d-1}}
x_{i_{1}\ldots i_{n}}\left\vert i_{1}\right\rangle \otimes\ldots
\otimes\left\vert i_{n}\right\rangle ,\ \ \ x_{i_{1}\ldots i_{n}}\in
\mathbb{C}, \label{qs-pn}%
\end{align}

\item with the normalization (\ref{qs-qu})%
\begin{equation}%
{\displaystyle\sum\limits_{i_{1}=0}^{d-1}}
\ldots%
{\displaystyle\sum\limits_{i_{n}=0}^{d-1}}
\left\vert x\right\vert _{i_{1}\ldots i_{n}}^{2}=1.
\end{equation}

\end{enumerate}
\end{definition}

\begin{definition}
The multi-superqudits ($n$-superqudit states) are

\begin{enumerate}
\item linear span of the super Hilbert subspace%
\begin{align}
\operatorname*{span}(\Vert\text{\textbf{\.{I}}}_{1}\rangle\otimes\ldots
\otimes\Vert\text{\textbf{\.{I}}}_{n}\rangle)  &  \subseteq\overset
{n}{\overbrace{\mathcal{H}_{\left(  r|s\right)  }\otimes\ldots\otimes
\mathcal{H}_{\left(  r|s\right)  }}},\ \ \ \Vert\text{\textbf{\.{I}}}%
_{k}\rangle=\left(  \Vert\mathbf{i}_{k}\rangle,\Vert\text{\textbf{\oe }}%
_{k}\rangle\right)  ,\nonumber\\
\Vert\mathbf{i}_{k}\rangle &  \in\mathcal{H}_{\left(  r|s\right)  }^{\left(
\bar{0}\right)  },\Vert\text{\textbf{\oe }}_{k}\rangle\in\mathcal{H}_{\left(
r|s\right)  }^{\left(  \bar{1}\right)  },\ \ \ k=1,\ldots,n,
\end{align}
which respect parities of variables, such that we have even and odd
superqubits (see(\ref{qs-vw}))%
\begin{align}
\Vert\mathbf{\Psi}^{\left(  \mathsf{k}\right)  }\left(  n\right)  \rangle &  =%
{\displaystyle\sum\limits_{j_{1}=0}^{n-1}}
\ldots%
{\displaystyle\sum\limits_{j_{n}=0}^{n-1}}
\mathbf{y}_{j_{1}\ldots j_{n}}\Vert\text{\textbf{\.{I}}}_{1}\rangle
\otimes\ldots\otimes\Vert\text{\textbf{\.{I}}}_{n}\rangle,\ \ \ \ \mathbf{y}%
_{j_{1}\ldots j_{n}}=\left(  \mathit{x}_{j_{1}\ldots j_{n}},\text{\ae }%
_{j_{1}\ldots j_{n}}\right)  ,\nonumber\\
\mathsf{k}  &  =\deg\mathbf{y}_{j_{1}\ldots j_{n}}\boxplus\deg\Vert
\text{\textbf{\.{I}}}_{1}\rangle\boxplus\ldots\deg\Vert\text{\textbf{\.{I}}%
}_{n}\rangle=\bar{0},\bar{1}\in\mathbb{Z}_{2}, \label{qs-pn1}%
\end{align}

\item normalization can be made for the even multi-superqubit $\Vert
\mathbf{\Psi}^{\left(  \bar{0}\right)  }\left(  n\right)  \rangle$ only, as
for the single superqubit (\ref{qs-qd1}).
\end{enumerate}
\end{definition}

To clarify the difference between the separable (\ref{qs-tp}) and nonseparable
(\ref{qs-pn}), (\ref{qs-pn1}) states we consider the example of two
(super)qubits. Thus, for $n=2$ (two-party states) we obtain%
\begin{align}
&  \left\vert \Psi\left(  2\right)  \right\rangle =x_{00}\left\vert
0\right\rangle \otimes\left\vert 0^{\prime}\right\rangle +x_{01}\left\vert
0\right\rangle \otimes\left\vert 1^{\prime}\right\rangle +x_{10}\left\vert
1\right\rangle \otimes\left\vert 0^{\prime}\right\rangle +x_{11}\left\vert
1\right\rangle \otimes\left\vert 1^{\prime}\right\rangle \nonumber\\
&  \Downarrow_{susy}\nonumber\\
&  \left(
\begin{array}
[c]{c}%
\Vert\mathbf{\Psi}^{\left(  \bar{0}\right)  }\left(  2\right)  \rangle\\
\Vert\mathbf{\Psi}^{\left(  \bar{1}\right)  }\left(  2\right)  \rangle
\end{array}
\right)  =\Vert\mathbf{0\rangle\otimes}\Vert\mathbf{0}^{\prime}\mathbf{\rangle
\left(
\begin{array}
[c]{c}%
\mathit{x}_{00}\\
\text{\ae }_{00}%
\end{array}
\right)  +}\Vert\mathbf{0\rangle\otimes}\Vert\mathbf{1}^{\prime}%
\mathbf{\rangle}\left(
\begin{array}
[c]{c}%
\mathit{x}_{01}\\
\text{\ae }_{01}%
\end{array}
\right) \nonumber\\
&  \mathbf{\mathbf{+}\Vert\mathbf{0\rangle\otimes}\Vert}\text{\textbf{\oe }%
}\mathbf{^{\prime}\mathbf{\rangle\left(
\begin{array}
[c]{c}%
\text{\ae }_{02}\\
\mathit{x}_{02}%
\end{array}
\right)  +}}\Vert\mathbf{1\rangle\otimes}\Vert\mathbf{0}^{\prime
}\mathbf{\rangle\left(
\begin{array}
[c]{c}%
\mathit{x}_{10}\\
\text{\ae }_{10}%
\end{array}
\right)  +}\Vert\mathbf{1\rangle\otimes}\Vert\mathbf{1}^{\prime}%
\mathbf{\rangle}\left(
\begin{array}
[c]{c}%
\mathit{x}_{11}\\
\text{\ae }_{11}%
\end{array}
\right) \nonumber\\
&  \mathbf{\mathbf{+}\Vert1\mathbf{\rangle\otimes}\Vert}\text{\textbf{\oe }%
}\mathbf{^{\prime}\mathbf{\rangle\left(
\begin{array}
[c]{c}%
\text{\ae }_{12}\\
\mathit{x}_{12}%
\end{array}
\right)  +}}\Vert\text{\textbf{\oe }}\mathbf{\rangle\otimes}\Vert
\mathbf{0}^{\prime}\mathbf{\rangle\left(
\begin{array}
[c]{c}%
\text{\ae }_{20}\\
\mathit{x}_{20}%
\end{array}
\right)  +}\Vert\text{\textbf{\oe }}\mathbf{\rangle\otimes}\Vert
\mathbf{1}^{\prime}\mathbf{\rangle}\left(
\begin{array}
[c]{c}%
\text{\ae }_{21}\\
\mathit{x}_{21}%
\end{array}
\right) \nonumber\\
&  \mathbf{+}\Vert\text{\textbf{\oe }}\mathbf{\rangle\otimes}\Vert
\text{\textbf{\oe }}^{\prime}\mathbf{\rangle}\left(
\begin{array}
[c]{c}%
\mathit{x}_{22}\\
\text{\ae }_{22}%
\end{array}
\right)  \mathbf{\in}\left(
\begin{array}
[c]{c}%
\left(  \mathcal{H}_{\left(  2|1\right)  }\otimes\mathcal{H}_{\left(
2|1\right)  }^{\prime}\right)  ^{\left(  \bar{0}\right)  }\\
\left(  \mathcal{H}_{\left(  2|1\right)  }\otimes\mathcal{H}_{\left(
2|1\right)  }^{\prime}\right)  ^{\left(  \bar{1}\right)  }%
\end{array}
\right)  , \label{qs-bi}%
\end{align}
where $\left\vert \Psi\left(  2\right)  \right\rangle $ has 4
\textquotedblleft bosons\textquotedblright, $\Vert\mathbf{\Psi}^{\left(
\bar{0}\right)  }\left(  2\right)  \rangle$ has 5 \textquotedblleft
bosons\textquotedblright\ and 4 \textquotedblleft fermions\textquotedblright,
$\Vert\mathbf{\Psi}^{\left(  \bar{1}\right)  }\left(  2\right)  \rangle$ has 4
\textquotedblleft bosons\textquotedblright\ and 5 \textquotedblleft
fermions\textquotedblright. Comparing the tensor product (\ref{qs-tp}) and
(\ref{qs-bi}), we observe that for separable states all the amplitudes
(\textquotedblleft coordinates\textquotedblright) in (\ref{qs-bi}) can be
composed%
\begin{equation}%
\begin{tabular}
[c]{lll}%
$\left.
\begin{array}
[c]{c}%
x_{ij}\overset{sep}{=}x_{i}x_{j}^{\prime},\\
x_{i}\in\mathbb{C},\\
i,j=0,1
\end{array}
\right.  $ & $\overset{susy}{\Longrightarrow}$ & $\left.
\begin{array}
[c]{c}%
\mathit{x}_{ij}\overset{sep}{=}\left\{
\begin{array}
[c]{c}%
\mathit{x}_{i}\mathit{x}_{j}^{\prime}\\
\text{\ae }_{i}\text{\ae }_{j}^{\prime}%
\end{array}
\right.  ,\mathit{x}_{i2}\overset{sep}{=}\left\{
\begin{array}
[c]{c}%
\mathit{x}_{i}\mathit{x}^{\prime}\\
\text{\ae }_{i}\text{\ae }^{\prime}%
\end{array}
\right.  ,\mathit{x}_{2i}\overset{sep}{=}\left\{
\begin{array}
[c]{c}%
\text{\ae \ae }_{i}^{\prime}\\
\mathit{xx}_{i}^{\prime}%
\end{array}
\right.  ,\mathit{x}_{22}\overset{sep}{=}\left\{
\begin{array}
[c]{c}%
\text{\ae \ae }^{\prime}\\
\mathit{xx}^{\prime}%
\end{array}
\right. \\
\text{\ae }_{ij}\overset{sep}{=}\left\{
\begin{array}
[c]{c}%
\mathit{x}_{i}\text{\ae }_{j}^{\prime}\\
\text{\ae }_{i}\mathit{x}_{j}^{\prime}%
\end{array}
\right.  ,\text{\ae }_{i2}\overset{sep}{=}\left\{
\begin{array}
[c]{c}%
\mathit{x}_{i}\text{\ae }^{\prime}\\
\text{\ae }_{i}\mathit{x}^{\prime}%
\end{array}
\right.  ,\text{\ae }_{2i}\overset{sep}{=}\left\{
\begin{array}
[c]{c}%
\text{\ae }\mathit{x}_{i}^{\prime}\\
\mathit{x}\text{\ae }_{i}^{\prime}%
\end{array}
\right.  ,\text{\ae }_{22}\overset{sep}{=}\left\{
\begin{array}
[c]{c}%
\text{\ae }\mathit{x}^{\prime}\\
\mathit{x}\text{\ae }^{\prime}%
\end{array}
\right. \\
\mathit{x},\mathit{x}_{i},\mathit{x}_{ij},\mathit{x}_{i2},\mathit{x}%
_{2i},\mathit{x}^{\prime},\mathit{x}_{i}^{\prime}\in\Lambda_{N}^{\left(
\bar{0}\right)  }\left(  \mathbb{C}\right)  ,\\
\text{\ae },\text{\ae }_{i},\text{\ae }_{ij},\text{\ae }_{i2},\text{\ae }%
_{2i},\text{\ae }^{\prime},\text{\ae }_{i}^{\prime}\in\Lambda_{N}^{\left(
\bar{1}\right)  }\left(  \mathbb{C}\right)  .
\end{array}
\right.  $%
\end{tabular}
\label{qs-x}%
\end{equation}

\begin{remark}
The separability of two-party superqubit even $\Vert\mathbf{\Psi}^{\left(
\bar{0}\right)  }\left(  2\right)  \rangle$ and odd $\Vert\mathbf{\Psi
}^{\left(  \bar{1}\right)  }\left(  2\right)  \rangle$ states (\ref{qs-bi}) is
determined in the nonunique way (\ref{qs-x}).
\end{remark}

\begin{definition}
Multi-(super)qubit states are called entangled (inseparable), if at least one
of their amplitudes ($\mathbf{y}_{j_{1}\ldots j_{n}}$ in (\ref{qs-pn1}))
cannot be presented in the composite factorized form (\ref{qs-x}).
\end{definition}

A suitable function which can measure entanglement should have the main
property: vanishing for the separable states (\ref{qs-tp}). The simplest such
function for two qubits (without other requirements) is the determinant.
Indeed, for the separable two party (super)qubit system we have from the
factorization (\ref{qs-x})%
\begin{align}
\left\vert \Psi\left(  2\right)  \right\rangle  &  :f\left(  x\right)  =\det
x_{ij}\overset{sep}{=}\det\left(  x_{i}x_{j}^{\prime}\right)  \equiv
0,\ \ \ \forall x_{i},x_{j}^{\prime}\in\mathbb{C},\ \ i,j=0,1.\label{qs-f}\\
&  \Downarrow_{susy}\nonumber\\
\Vert\mathbf{\Psi}^{\left(  \bar{0}\right)  }\left(  2\right)  \rangle &
:\mathit{f}^{\left(  \bar{0}\right)  }\left(  \mathit{y}\right)  =\det\left(
\mathit{x}_{ij}\mathit{x}_{22}+\text{\ae }_{i2}\text{\ae }_{2j}\right)
\overset{sep}{=}0,\ \ \mathit{y}_{ij}=\left(  \mathit{x}_{ij},\text{\ae }%
_{ij}\right)  ,\label{qs-f0}\\
\Vert\mathbf{\Psi}^{\left(  \bar{1}\right)  }\left(  2\right)  \rangle &
:\mathit{f}^{\left(  \bar{1}\right)  }\left(  \mathit{y}\right)  =\det\left(
\text{\ae }_{ij}\text{\ae }_{22}-\mathit{x}_{i2}\mathit{x}_{2j}\right)
\overset{sep}{=}0,\mathit{x}_{ij}\in\Lambda_{N}^{\left(  \bar{0}\right)
}\left(  \mathbb{C}\right)  ,\text{\ae }_{ij}\in\Lambda_{N}^{\left(  \bar
{1}\right)  }\left(  \mathbb{C}\right)  . \label{qs-f1}%
\end{align}

Further requirements can be imposed, for ordinary qubits they are positivity,
monotonicity and the range in $\left\{  0,1\right\}  $, as probability, which
gives the concurrence \cite{hil/woo,woo98,hor2009}%
\begin{equation}
C_{2}\left(  x\right)  =C\left(  \left\vert \Psi\left(  2\right)
\right\rangle \right)  =2\left\vert f\left(  x\right)  \right\vert
=2\left\vert \det x_{ij}\right\vert , \label{qs-cx}%
\end{equation}
such that for maximally entangled states, e.g. the Bell state $x_{11}%
=x_{22}=\frac{1}{\sqrt{2}}$, $x_{01}=x_{10}=0$, to get $C\left(  x\right)  =1$.

We define the even and odd superconcurrences by%
\begin{align}
\mathit{C}^{\left(  \bar{0}\right)  }\left(  \mathit{y}\right)   &
=\mathit{C}\left(  \Vert\mathbf{\Psi}^{\left(  \bar{0}\right)  }\left(
2\right)  \rangle\right)  =2\left\Vert \det\left(  \mathit{x}_{ij}%
\mathit{x}_{22}+\text{\ae }_{i2}\text{\ae }_{2j}\right)  \right\Vert
_{R},\label{qs-c0}\\
\mathit{C}^{\left(  \bar{1}\right)  }\left(  \mathit{y}\right)   &
=\mathit{C}\left(  \Vert\mathbf{\Psi}^{\left(  \bar{1}\right)  }\left(
2\right)  \rangle\right)  =2\left\Vert \det\left(  \text{\ae }_{ij}%
\text{\ae }_{22}-\mathit{x}_{i2}\mathit{x}_{2j}\right)  \right\Vert
_{R},\label{qs-c1}%
\end{align}
where $\left\Vert \_\right\Vert _{R}$ is one of the Grassmann norms
\cite{dewitt,rogers,rud2000}.

The square of the concurrence is called tangle \cite{bor/dah/duf/rub}, which
can be written for two qubits in the form%
\begin{equation}
\tau\left(  x\right)  =4f\left(  x\right)  \overline{f\left(  x\right)
}=4\det x_{ij}\det\bar{x}_{ij}, \label{qs-tx}%
\end{equation}
where $\overline{\left(  \_\right)  }$ is the complex conjugation.

For two even/odd superqubits, by analogy with (\ref{qs-tx}) and taking into
account possible noninvertibilities we can define the even supertangle
$\mathbf{\tau}^{\left(  \bar{0}\right)  }\left(  x\right)  $ and odd
supertangle $\mathbf{\tau}^{\left(  \bar{1}\right)  }\left(  x\right)  $ in
the following way%
\begin{align}
\mathbf{\tau}^{\left(  \bar{0}\right)  }\left(  \mathit{y}\right)
\mathit{x}_{22}\left(  \mathit{x}_{22}\right)  ^{\sharp} &  =4\mathit{f}%
^{\left(  \bar{0}\right)  }\left(  \mathit{y}\right)  \left(  \mathit{f}%
^{\left(  \bar{0}\right)  }\left(  \mathit{y}\right)  \right)  ^{\sharp
},\label{qs-t1}\\
\mathbf{\tau}^{\left(  \bar{1}\right)  }\left(  \mathit{y}\right)
\text{\ae }_{22}\left(  \text{\ae }_{22}\right)  ^{\sharp} &  =4\mathit{f}%
^{\left(  \bar{1}\right)  }\left(  \mathit{y}\right)  \left(  \mathit{f}%
^{\left(  \bar{1}\right)  }\left(  \mathit{y}\right)  \right)  ^{\sharp
},\label{qs-t2}%
\end{align}
where $\mathit{f}^{\left(  \bar{0}\right)  }\left(  \mathit{y}\right)  $ and
$\mathit{f}^{\left(  \bar{1}\right)  }\left(  \mathit{y}\right)  $ are defined
in (\ref{qs-f0}) and (\ref{qs-f1}), respectively.

In case of invertible $\mathit{x}_{22}$ the even superconcurrence
$\mathit{C}^{\left(  \bar{0}\right)  }\left(  \mathit{y}\right)  $
(\ref{qs-c0}) and even supertangle $\mathbf{\tau}^{\left(  \bar{0}\right)
}\left(  x\right)  $ (\ref{qs-t1}) can be connected with the Berezinian
(\ref{qs-ber}).

There are many other entanglement measures: entropy of entanglement, positive
partial transpose, quantum discord, entanglement of formation, distillable
entanglement, entanglement cost, squashed entanglement, entanglement witnesses
\cite{hor2009}. Some of them can be applied for multi-(super)qudits as well,
for superqubits, see, e.g. \cite{bor/dah/duf/rub}.

The entanglement classification and manipulation can be provided by
considering various local symmetries of multi-(super)qubit systems. The main
paradigm is Local Operations and Classical Communication (LOCC) proposed in
\cite{ben/div/smo}: it is not possible to change quantum property of a many
party state (e.g. increase its entanglement) using local operations (e.g. on
one party qubits) and classical channels only. Thus, the many party quantum
states can be classified in such a way, that each class contains the
represenative state with maximum entanglement. If some operations can be
performed using LOCC, but may fail, they are called Stochastic Local
Operations and Classical Communication (SLOCC) \cite{vid2000,ver/deh/dem}.
Quantum states which can be transformed into one another are called SLOCC
equivalent and the corresponding equivalence classes are called entanglement
(SLOCC) classes which are invariant under invertible unitary transformations
\cite{elt/sie,ver/deh/dem}.

A single qubit $\left\vert \Psi\right\rangle $ (\ref{qs-qd1}) carries the
fundamental representation of $SU\left(  2\right)  $ group, and therefore for
$n$-qubit state the LOCC equivalence group is $\left[  SU\left(
2,\mathbb{C}\right)  \right]  ^{\otimes n}$ \cite{vid2000,ver/deh/dem}, while
the SLOCC equivalence group is $\left[  SL\left(  2,\mathbb{C}\right)
\right]  ^{\otimes n}$ \cite{bor/dah/duf/rub}. Thus, any separable $n$-qubit
state will remain separable under all $\left[  SU\left(  2\right)  \right]
^{\otimes n}$ operations. In a similar way, the even superqubit $\Vert
\mathbf{\Psi}^{\left(  \bar{0}\right)  }\rangle$ (\ref{qs-qd1}) carries the
fundamental representation of the local operation (unitary orthosymplectic)
group $uOSp\left(  2|1\right)  $, and so for $n$-superqubit state the LOCC
equivalence group is $\left[  uOSp\left(  2|1\right)  \right]  ^{\otimes n}$
and the SLOCC equivalence group is $\left[  OSp\left(  2|1\right)  \right]
^{\otimes n}$ \cite{bor/bra/duf1}.

The supersymmetrization of (S)LOCC groups is different from
supersymmetrization of the Poincar\'{e} group, and therefore adding
arttificially superpartners of electron and photon does not give a superqubit
\cite{bra2012}. Nevetheless, supersymmetric extension of quantum mechanics
based on superqubits may be a candidate for a superquantum theory that lies in
the gap between the ordinary quantum theory and non-local boxes
\cite{pop/roh,bor/bra/duf}. There can be applications of superqubits in
condensed matter physics where the orthosymplectic Lie superalgebras play an
important role \cite{efetov}.